\documentclass[aip,jcp,sd,amsmath,amssymb,reprint,%
]{revtex4-1}
\usepackage{natbib}
\setcitestyle{super}

\newcommand*{\citen}[1]{%
  \begingroup
    \romannumeral-`\x % remove space at the beginning of \setcitestyle
    \setcitestyle{numbers}%
    \cite{#1}%
  \endgroup   
}

\usepackage{float,graphicx,subcaption}% Include figure files
\usepackage{color}
\usepackage{dcolumn}% Align table columns on decimal point
\usepackage{bm}% bold math
\newcommand{\moy}[1]{\left\langle #1 \right\rangle}

\newcommand{\XX}[0]{\boldsymbol{X}}

\newcommand{\ex}[1]{\mathrm{e}^{#1}}

\newcommand{\dd}[0]{\mathrm{d}}

\newcommand{\rr}[0]{\boldsymbol{r}}
\newcommand{\RR}[0]{\boldsymbol{R}}

\newcommand{\pp}[0]{\boldsymbol{p}}

\newcommand{\kB}[0]{k_{\mathrm{B}}}
\newcommand{\LL}[0]{\boldsymbol{l}}

\newcommand{\nn}[0]{\hat{\boldsymbol{n}}}

\newcommand{\rotop}[0]{\boldsymbol{\mathcal{R}}}
\newcommand{\rotopt}[0]{\tilde{\boldsymbol{\mathcal{R}}}}
\newcommand{\MM}[0]{\mathbf{M}}
\newcommand{\VV}[0]{\mathbf{V}}
\newcommand{\WW}[0]{\mathbf{W}}

\newcommand{\muO}[0]{\boldsymbol{\mu}^{11}}
\newcommand{\muT}[0]{\boldsymbol{\mu}^{22}}
\newcommand{\muOT}[0]{\boldsymbol{\mu}^{12}}

\newcommand{\II}[0]{\boldsymbol{1}}
\newcommand{\sgm}[0]{\boldsymbol{\sigma}}
\newcommand{\AAA}[0]{\boldsymbol{A}}

\newcommand{\uth}[0]{\boldsymbol{\hat{\theta}}}
\newcommand{\uph}[0]{\boldsymbol{\hat{\varphi}}}
\newcommand{\DD}[0]{\boldsymbol{D}}
		
\begin{document}

\preprint{AIP/123-QED}

\title[Chemical and hydrodynamic alignment of an enzyme]{Chemical and hydrodynamic alignment of an enzyme}

\author{T. Adeleke-Larodo}
\email{tunrayo.adeleke-larodo@physics.ox.ac.uk.}
\affiliation{Rudolf Peierls Centre for Theoretical Physics, University of Oxford, Oxford OX1 3PU, United Kingdom}

\author{J. Agudo-Canalejo}%
\email{jaime.agudocanalejo@physics.ox.ac.uk.}
\affiliation{Rudolf Peierls Centre for Theoretical Physics, University of Oxford, Oxford OX1 3PU, United Kingdom}
\affiliation{Department of Chemistry, The Pennsylvania State University, University Park, Pennsylvania 16802, United States}
\author{R. Golestanian}
\email{ramin.golestanian@ds.mpg.de}
\affiliation{Max Planck Institute for Dynamics and Self-Organization (MPIDS), Am Fassberg 17, D-37077 G${\ddot{\rm o}}$ttingen, Germany}
\affiliation{Rudolf Peierls Centre for Theoretical Physics, University of Oxford, Oxford OX1 3PU, United Kingdom}

\date{\today}% It is always \today, today,

\begin{abstract}
Motivated by the implications of the complex and dynamic modular geometry of an enzyme on its motion, we investigate the effect of combining long-range internal and external hydrodynamic interactions due to thermal fluctuations with short-range surface interactions. An asymmetric dumbbell consisting of two unequal subunits, in a nonuniform suspension of a solute with which it interacts via hydrodynamic interactions as well as non-contact surface interactions, is shown to have two alignment mechanisms due to the two types of interactions. In addition to alignment, the chemical gradient results in a drift velocity that is modified by hydrodynamic interactions between the constituents of the enzyme. 
\end{abstract}

\maketitle
\section{\label{sec:Intro}Introduction}
While the role of enzymes in accelerating and regulating life-sustaining biochemical reactions inside cells is widely accepted, the mechanical response of an enzyme to the biochemical environment and its effect on how the enzyme functions has recently become a topic of interest because of the potential implications in understanding metabolic processes \cite{sweetlove_2018,Wu_2015} and the possiblity of exploiting enzyme functionalities to produce biocompatible micro- and nanoscale controllable machines \cite{sengupta_2013,Dey_2015}. In the context of self-propelled low-Reynolds number particles, propulsion due to phoretic effects has been studied extensively, theoretically and experimentally \cite{illien_2017_CSR}. The question that is now being asked is whether enzyme molecules, undergoing catalytic turnover in varying reactant concentrations can exhibit similar effects.

A number of experiments have reported enhanced diffusion of an appreciable number of different enzymes in the presence of a homogeneous distribution of their substrate, with a Michaelis-Menten dependance of the diffusion coefficient on substrate concentration \cite{yu_2009,muddana_2010,sengupta_2013,sengupta_2014,Riedel_2015,Illien_2017_NL}. More recently there have been experimental reports of directed motion of catalytically active enzymes when the substrate concentration is nonuniform so that there is a concentration gradient. The phenomenon has been observed for various enzymes and has typically been seen to cause movement  towards the concentration gradient (reported in catalase and urease \cite{sengupta_2013,dey_2014}, RNA Polymerase \cite{yu_2009}, DNA Polymerase \cite{sengupta_2014}, and hexokinase and aldolase \cite{zhao_2018}). The opposite has also been reported, where the enzymes (urease and acetylcholinesterase) are seen to move away from their substrate \cite{jee_2014}. These observations are surprising because of the size of a single enzyme molecule, and its implication on the necessary time-scale of such a dynamics. A perceivable response of an enzyme to a local gradient would be the culmination of overcoming thermal fluctuations and viscous effects that are expected to have significant effects on dynamics at the micro- and nanoscopic length-scales, and are amplified inside a cell. Several possible theories have been suggested that partially address the experimental observations \cite{Agudo-Canalejo_2018_ACR}.

Initial theories to explain enhanced diffusion of enzymes relied on nonequilibrium aspects of the catalytic cycle \cite{Golestanian_2010,Riedel_2015,Golestanian_2015,Bai_2015,Hwang_2017}. We recently proposed an asymmetric dumbbell model for an enzyme to study the diffusion of a single molecule \cite{illien_2017_EPL}. With this model we proposed an equilibirum mechanism for the phenomenon of enhanced diffusion of an enzyme in the presence of its substrate, or indeed any molecule that is able to occupy the binding site and induce conformational changes in the enzyme, such as an inhibitor. A natural question is to ask whether these effects are also relevant to the motion of an enzyme in the presence of concentration gradients of their substrate.

In this paper the specific geometry of an asymmetric dumbbell is adopted to study the response of the modular structure of an oligomeric enzyme to an externally imposed gradient of a chemical that interacts with the enzyme, and is able to occupy the binding site. We find that coupling solvent-mediated hydrodynamic interactions between the subunits of our model enzyme and the molecules of the chemical field, and the non-covalent interactions with the chemical molecules leads to alignment of the enzyme by two mechanisms. In a fluid containing substrate $\text{s}$ with concentration $\rho_\text{s}$, in the absence of substrate binding, the probability density $\tilde{\rho}_\text{e}(\RR, \nn; t)$ of the model enzyme being located at position $\RR$ with orientation $\nn$ evolves under the following Smoluchowki equation:
\begin{eqnarray}
\label{eq:canon}
\partial_t\tilde{\rho}_\text{e}(\RR, \nn; t) &&= \nabla_{\RR} \cdot [ \DD^t \cdot \nabla_{\RR} \tilde{\rho}_\text{e} - (\boldsymbol{\mu}^\text{v}\cdot\nabla_{\RR} \rho_\text{s}) \tilde{\rho}_\text{e} ]\\
&& + \rotop \cdot [ D^r \rotop \tilde{\rho}_\text{e} - \mu^\omega( \nn \times \nabla_{\RR} \rho_\text{s}) \tilde{\rho}_\text{e}]\nonumber\\
&& + \rotop \cdot [ \DD^c \cdot \nabla_{\RR} \tilde{\rho}_\text{e} ]+ \nabla_{\RR} \cdot [ (\DD^c)^\mathrm{T} \cdot \rotop \tilde{\rho}_\text{e}]\nonumber.
\end{eqnarray}
The first line has a generic form, describing motion that is not specific to the dumbbell geometry: Translational diffusion with diffusion coefficient $\DD^t$ and diffusiophoretic drift with velocity $\VV_\text{ph}=\boldsymbol{\mu}^\text{v}\cdot\nabla_{\RR} \rho_\text{s}$ due to non-covanlent interactions of the substrate with the surface of the enzyme \cite{anderson_1989,stone_1996}. For the dumbbell, the phoretic mobility $\boldsymbol{\mu}^\text{v}$ is the sum of the phoretic mobilities of the individual subunits due to diffusiophoresis and a correction due to coupling of the subunits. The second line characterises the rotational motion of the enzyme: Rotational diffusion of the orientation vector $\nn$ with diffusion coefficient $D^r$ and a term corresponding to alignment of the orientation vector parallel or anti-parallel to the concentration gradient that is controlled by $\mu^\omega$. This alignment mechanism was previously known \cite{saha_2014} and achieved by artificial symmetry breaking (by methods such as patterning the surface of a spherical particle with a substance that undergoes catalytic activity with the chemical in the bulk), here emerges naturally as a result of the in-built asymmetry of an enzyme. The final line of Eq.~(\ref{eq:canon}) contains contributions to the motion due to the asymmetric dumbbell geometry, where $\DD^c$ is a tensor coupling translational and rotational motion of the subunits and $(\DD^c)^\mathrm{T}$ is the transpose tensor. The first term gives a purely hydrodynamic contribution to alignment that comes from gradients in the density field.

It has been shown that the action of binding of a substrate molecule to an enzyme to form an enzyme-substrate complex results in an additional contribution to the drift velocity of an enzyme in a substrate concentration gradient, due to the difference in diffusivity between the free and bound states \cite{agudo-Canalejo_2018}. Furthermore, that the drift velocity, in addition to the diffusion coefficient previously reported in experiments and theory, is a Michaelis-Menten average over the two enzyme states.  A similar consideration of a reduced enzyme kinetics for our system reveals a universal modification of the transport properties by the fluctuation-induced hydrodynamic interactions. When the mean time for formation of an enzyme-substrate complex is much greater than the rotational diffusion time of the enzyme (this could be an attribute of a specific enzyme, or because the solution is sufficiently dilute), the evolution equation can be written in the form
\begin{eqnarray}
\label{eq:c_tot}
\partial_t c_\text{tot}(\RR;t) &&= \nabla_{\RR}\cdot\Big\{D^\text{eff}(\RR)\nabla_{\RR}c_\text{tot}\nonumber\\
&& - \left[\VV^\text{eff}_\text{ph}(\RR)+\VV^\text{eff}_\text{bi}(\RR)\right]c_\text{tot}\Big\},
\end{eqnarray}
with an effective diffusion coefficient $D^\text{eff}$ and velocities $\VV^\text{eff}_\text{ph}$ and $\VV^\text{eff}_\text{bi}$ due to phoresis and binding interactions.

In the following section we describe the system and present the main steps in our calculation for determining the alignment mechanisms for our model enzyme in an interacting chemical field. Specific details of the calculation are reserved for the supporting appendices. Following this, an effective mobility is derived for the long-time, large length-scale dynamics. The alignment of the enzyme due to diffusiophoresis is examined in more detail. In the final results section, we consider the effect of the binding and unbinding of substrate molecules on the directed motion of our enzyme. Finally, we conclude with a discussion of the significance and possible implications of our results for further investigations into the dynamics of enzymes, and other low-Reynolds number modular structures for which internal hydrodynamics may play a role.
\section{\label{model}Model}
\begin{figure*}[t]
  \centering
  \subcaptionbox{}[.3\linewidth][c]{%
    \includegraphics[width=.25\linewidth]{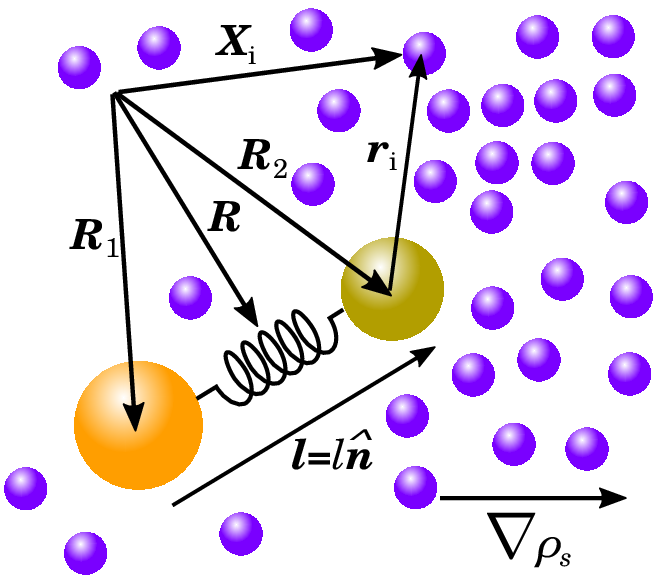}}\quad
  \subcaptionbox{}[.6\linewidth][c]{%
\includegraphics[width=0.6\textwidth]{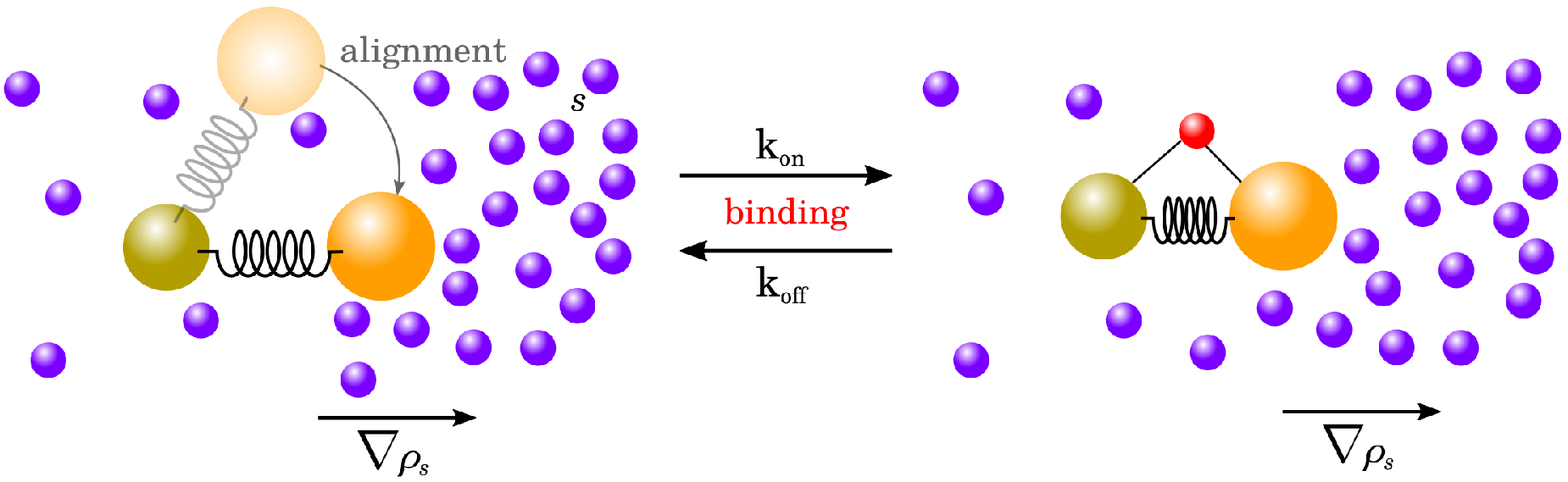}}\quad
  \caption{(a) A free enzyme in a gradient of substrate molecules. The free enzyme interacts with substrate molecules in the bulk via pairwise hydrodynamic and non-covalent surface interactions. (b) Additionally, the free enzyme can bind to a substrate molecule to form a complex with rate $k_\text{on}$ which decomposes at rate $k_\text{off}$.}
  \label{fig:dumbbell}
\end{figure*}
For many enzymes there is a separation of timescales between their dynamics and reactions that allows us to make simplifying approximations. Consider the following simple model for enzyme catalysis, where substrate approach the enzyme by a diffusive process: $\text{E} + \text{\text{S}} \leftrightarrows\text{ES}\leftrightarrows\text{E}+\text{P}$.
The characteristic timescales of the system are identified as the following: The mean binding time for the formation of an enzyme-substrate complex $\text{ES}$; the mean time for catalysis, when product are formed; the relaxation time of internal fluctuations of the enzyme; the translational and rotational diffusion times; and the relaxation time of the fluid around the enzyme \cite{lee_1987}. Initially, the reactions of the enzyme are ignored with the understanding that the relaxation times for the internal dynamics, and the diffusion times for both the enzyme and substrate molecules are much smaller. For most enzymes, the catalytic step is typically orders of magnitude slower than the diffusion of substrates and products.

Submerged in an inhomogeneous substrate concentration, we consider an asymmetric dumbbell consisting of two unequal subunits, taken to be rigid spheres to simplify the hydrodynamic treatment, which are subjected to thermal fluctuations that mediate hydrodynamic interactions between the subunits and between the subunits and substrate molecules in the bulk. In addition to the solvent-mediated interactions, there is an interaction potential between the subunits which represents the non-covalent interactions between the parts of an oligomeric  enzyme. A schematic of the set-up is shown in Fig.~\ref{fig:dumbbell}. Indeed, the majority of known enzymes are oligomeric. Each subunit has separate direct interactions with the substrate molecules in the bulk, which can be either attractive or repulsive, for example, coming from van der Waals, electrostatic or steric forces. 
\section{\label{alignment}Single molecule alignment}
The stochastic dynamics of the model enzyme, with subunits at positions $\RR_1$ and $\RR_2$, in an unbounded bath of $N$ substrate molecules at positions $\XX_1,\dots,\XX_N$, is characterized by the $(N+2)$-particle distribution $\rho(\RR_1,\RR_2,\XX_1,\dots,\XX_N;t)$. We assume that the distribution evolves under the Smoluchowski equation
\begin{eqnarray}
&&\partial_t\rho(\RR_1,\RR_2,\XX_1,\dots,\XX_N;t)=\nonumber\\
&& \sum_{i,j=1}^2 \nabla_{\RR_i} \cdot \left( \boldsymbol{\mu}^{ij}\cdot[\kB T \nabla_{\RR_j}\rho+  (\nabla_{\RR_j}\phi)\rho] \right)\nonumber\\
&& +\sum_{i=1}^N \Big\{ \sum_{j=1}^2  \Big[ \nabla_{\RR_j} \cdot \left( \boldsymbol{\mu}^{js}\cdot[\kB T \nabla_{\XX_i}\rho+ (\nabla_{\XX_i}\phi)\rho] \right) \nonumber\\
&&+ \nabla_{\XX_i} \cdot \left( \boldsymbol{\mu}^{sj}\cdot[\kB T \nabla_{\RR_j}\rho+  (\nabla_{\RR_j}\phi)\rho] \right) \Big]\nonumber\\
&&+\nabla_{\XX_i} \cdot \left(\boldsymbol{\mu}^{ss}\cdot[\kB T \nabla_{\XX_i}\rho+ (\nabla_{\XX_i}\phi)\rho] \right) \Big\},
\label{eq:smol_1}
\end{eqnarray}
where $\boldsymbol{\mu}^{\alpha\beta}$ are the hydrodynamic mobility tensors. Strictly, the $\boldsymbol{\mu}^{\alpha\beta}$ as well as the interaction potential $\phi$ depend on the position of all $(N+2)$ particles. In a sufficiently dilute solution, the interaction potential $\phi$ can be approximated as the sum of pair potentials between the dumbbell subunits and the pair potential between subunit $j$ and a substrate molecule, 
$\phi(\RR_1,\RR_2,\XX_1,\dots,\XX_N)=\phi^\text{e}(\RR_1,\RR_2) + \sum_{j=1}^2 \sum_{i=1}^N \phi^{js}(\RR_j,\XX_i)
$, in addition if the substrate molecules are small, we can assume that the subunit self-mobilities $\mu^{ii}$ are constant, and for $i \neq j$, the cross-mobilities $\mu^{ij}$ contain only the pair interactions. In its current form Eq.~(\ref{eq:smol_1}) holds information on the evolution of the model enzyme as well as the $N$ substrate molecules. Since we are only interested in the evolution of the enzyme, we define the two-particle distribution
\begin{eqnarray}
\rho_\text{e}(\RR_1,\RR_2;t)&&=\\
&& \int \dd \XX_1 \dots \dd \XX_N\,  \rho(\RR_1,\RR_2,\XX_1,\dots,\XX_N)\nonumber.
\label{eq:2-particle}
\end{eqnarray}
The evolution equation for the two-particle distribution is not closed as it retains information of higher order interactions through the three-particle distribution $\rho_\text{es}(\RR_1,\RR_2, \XX;t)$. This is the first of the Bogoliubov-Born-Green-Kirkwood-Yvon-type (BBGKY) hierarchies we will face. Here the evolution of $\rho_\text{e}$ and $\rho_\text{es}$ are coupled through solvent-mediated hydrodynamic interactions and the interaction potential $\phi^{jS}$. For a dilute solution, a natural closure approximation is 
\begin{equation}
\rho_\text{es}(\RR_1,\RR_2,\XX;t) \simeq \rho_\text{e}(\RR_1,\RR_2) \frac{\rho_\text{s}(\XX;t)}{N} \ex{-\frac{\phi^{1s}+\phi^{2s}}{\kB T}}.
\label{eq:dbclosure}
\end{equation}
A second approximation is needed in order to be able to write the remaining integral over the substrate position into a manifestly phoretic form. In Eq.~(\ref{eq:dbclosure}) there is still a coupling of the interactions $\phi^{1s}$ and $\phi^{2s}$ between the subunits and the substrate molecules through the Boltzmann weight. We assume that the range of the pair potentials is much shorter than the typical distance between subunits $1$ and $2$, that is, that the two potentials do not overlap, implying that a substrate particle never `feels' simultaneously both subunits at the same time. This is equivalent to the assumption that the subunits are much larger than a substrate molecule and separated by a distance that is also larger than the substrate size. The evolution equation for the two-particle distribution (\ref{eq:2-particle}) can then be written as
\begin{eqnarray}
&&\partial_t \rho_\text{e}(\RR_1,\RR_2;t) =\nonumber\\
\label{eq:twopart2}
&& \sum_{i,j=1}^2 \Bigg\{ \nabla_{\RR_i} \cdot \left( \boldsymbol{\mu}^{ij} \cdot  \left[ \kB T \nabla_{\RR_j}\rho_\text{e} + (\nabla_{\RR_j} \phi^\text{e})\rho_\text{e}   \right] \right)\\
&& + \nabla_{\RR_i} \cdot \left[\kB T \rho_\text{e} \int_{\XX}(\boldsymbol{\mu}^{is} - \boldsymbol{\mu}^{ij}) (\ex{-\frac{\phi^{js}}{\kB T}} - 1) \nabla_{\XX} \rho_\text{s} \right] \Bigg\}\nonumber.
\end{eqnarray}
Alone, the second line describes the evolution of the enzyme in a fluid in the absence of a substrate concentration gradient. The third line corresponds to phoretic response to the chemical field: For $i=j$ it describes diffusiophoresis of a single particle, and for $i\neq j$, the term is unique to the case of a dumbbell, given that it mixes the non-specific direct interactions of the substrate particles with the $j$ particle with the hydrodynamic interactions between  substrate and the $i$ particle, and the hydrodynamic interactions between $i$ and $j$ (i.e. 1 and 2).

We evaluate the integral in (\ref{eq:twopart2}) representing the phoretic response of the enzyme by assuming that the substrate particles are point-like; that the interaction $\phi$ is very short-ranged, and that the concentration $\rho_\text{s}$ in the proximity of a subunit is the solution to the Laplace equation with no flux boundary conditions at the surface of the subunit when there is a uniform gradient $\nabla \rho_\text{s}^\infty$ at infinity. To the order considered, the integrals corresponding to single-particle phoresis have additional contributions which come from solving for the substrate concentration field in the presence of both subunits simultaneously. Introducing the results for the integration of the phoretic terms into (\ref{eq:twopart2}), and defining the centre and elongation coordinates of the dumbbell, $\RR = (\RR_1 + \RR_2)/2$ and $\LL = \RR_2 - \RR_1$, with $\LL = l \nn$, the Smoluchowski equation can be rewritten as
\begin{eqnarray}
&&\partial_t\rho_\text{e}(\RR, \boldsymbol{l}; t) =\nonumber\\
&&  + \frac{1}{2}\nabla_{\RR}\cdot\left[\frac{1}{2}\MM\cdot\kB T\nabla_{\RR}\rho_\text{e}+\boldsymbol{\Gamma}\cdot(\kB T\nabla_{\boldsymbol{l}}\rho_\text{e}+(\nabla_{\boldsymbol{l}}U)\rho_\text{e})\right]\nonumber\\
&&+\nabla_{\boldsymbol{l}}\cdot\left[\WW\cdot(\kB T\nabla_{\boldsymbol{l}}\rho_\text{e}+(\nabla_{\boldsymbol{l}}U)\rho_\text{e})+\frac{1}{2}\boldsymbol{\Gamma}\cdot\kB T\nabla_{\RR}\rho_\text{e}\right]\nonumber\\
&&-\nabla_{\RR} \cdot \left\{ \frac{\kB T}{2\eta} \rho_\text{e} (\sgm_1+\sgm_2) \cdot \nabla_{\RR} \rho_\text{s}^\infty \right\}\nonumber\\
&&- \nabla_{\LL} \cdot \left\{ \frac{\kB T}{\eta} \rho_\text{e}(\sgm_2-\sgm_1) \cdot \nabla_{\RR} \rho_\text{s}^\infty \right\}
\label{eq:smolcom}
\end{eqnarray}
where the translation, rotation and coupling tensors are defined respectively as the following linear combinations of the hydrodynamic mobility tensors
\begin{eqnarray}
&&\MM =\muO + \muT + 2\muOT,\nonumber\\
&&\WW = \muO + \muT - 2\muOT,\nonumber\\
&&\boldsymbol{\Gamma} = \muT-\muO,
\end{eqnarray}
we have performed the relabelling $\phi^\text{e} (\RR_1,\RR_2) = U(l)$, and have defined
\begin{equation}
\sgm_1 \equiv  A_1\boldsymbol{1} + \frac{a_2^3}{l^3} \left( B_2 - \frac{3}{2} A_1 \right) \left(  \nn \nn - \frac{\boldsymbol{1}}{3} \right)
\end{equation}
and
\begin{equation}
\sgm_2 \equiv  A_2\boldsymbol{1} + \frac{a_1^3}{l^3} \left( B_1 - \frac{3}{2} A_2 \right) \left(  \nn \nn - \frac{\boldsymbol{1}}{3} \right).
\end{equation}
The $a_i$ are the sizes of the subunits and $A_i$ and $B_i$ are the values of the $\XX$ integral in (\ref{eq:twopart2}) for $i=j$ and $i\neq j$, respectively. For very short-ranged interactions we can use $r_i = a_i + \delta$ with $\delta \ll a_i$ for the distance between subunit $i$ and the substrate molecule at position $\XX$, giving to the lowest order
\begin{equation}
A_i \approx \int_0^\infty \mathrm{d}\delta \delta ( \ex{-\frac{\phi^{is}(\delta)}{\kB T}} - 1) \equiv  \lambda_i^2,
\label{eq:Ai}
\end{equation}
where $\lambda_i$ is the Derjaguin length and
\begin{equation}
B_i \approx \frac{a_i}{10} \int^\infty_0 \text{d}\delta\,(\ex{-\frac{\phi^{is}(\delta)}{\kB T}}-1) \equiv \frac{a_i}{10} \gamma_i,
\label{eq:Bi}
\end{equation}
where we have defined $\gamma_i$, which is a lengthscale of the order of the interaction range, but distinct from the Derjaguin length. Comparison of (\ref{eq:Ai}) and (\ref{eq:Bi}) shows that generally we should expect $B_i \gg A_i$, given that $A_i$ is of the order of the interaction range squared, whereas $B_i$ is of the order of the particle size times the interaction range. Typically the Derjaguin length is of the order of a few angstroms \cite{anderson_1989} and the hydrodynamic radius of an enzyme is of the order of nm, about 7 nm for urease \cite{muddana_2010}.

Eq.~(\ref{eq:smolcom}) describes the probabilistic evolution in the complete phase space of a single enzyme, at a time that is less than the mean association time for the formation of an enzyme-substrate complex. This is a consequence of the assumption that the solution is dilute. However, it still retains information of the slow ($\RR$) and fast ($\LL$) dynamics. The fast dynamics is composed of fluctuations around the equilibrium separation of the subunits that effect the confirmations of the enzyme, and diffusion of the orientation vector $\nn$. Eq.~(\ref{eq:smolcom}) is simplified by a reduction of the phase space to include only the relevant degrees of freedom, here, $\nn$ and $\RR$. The relaxation time for vibrations in $l$ is smaller than the rotational diffusion time because the ratio between them is the relative deformation of the enzyme due to thermal fluctuations, and is thus less than one \cite{illien_2017_EPL}, therefore, the dependance on $l$ is eliminated by considering sufficiently large times. Explicitly, assuming that the $l$-dependence of $\rho_\text{e}$ is Boltzmann-like, if we define the separation-averaged two-particle distribution $\tilde{\rho}_\text{e}=\int \text{d}l\, l^2 \rho_\text{e}$, and the average of a function that depends on the subunit separation $\moy{f}=\frac{1}{\tilde{\rho}_\text{e}}\int\text{d}l\, l^2\,f(l)\,\rho_\text{e}$, averaging (\ref{eq:smolcom}) in this way while treating the orientation as a constant, we arrive at (\ref{eq:canon}), with translational diffusion tensor
\begin{equation}
\DD^t \equiv \frac{1}{4} \kB T \moy{\MM},
\end{equation}
rotational diffusion coefficient
\begin{equation}
D^r \equiv  \kB T \moy{\frac{W_I}{l^2}},
\end{equation}
translation-rotation coupling tensor
\begin{equation}
[\DD^c]_{ij} \equiv \frac{\kB T }{2}\moy{\frac{\Gamma_I}{l}} \epsilon_{ikj} \hat{n}_k,
\end{equation}
with transpose $(\DD^c)^\mathrm{T}$, satisfiying $(\DD^c)^\mathrm{T}= - \DD^c$, phoretic mobility tensor
\begin{eqnarray}
&&\boldsymbol{\mu}^\text{v} \equiv \frac{\kB T}{2\eta} \Bigg[ (A_1 + A_2)\boldsymbol{1}+\\
&& \moy{\frac{1}{l^3}} \left(a_1^3 B_1 + a_2^3 B_2 - \frac{3}{2} (a_2^3 A_1 + a_1^3 A_2 ) \right) \left(\nn \nn - \frac{\boldsymbol{1}}{3} \right) \Bigg]\nonumber\\
&&= \left[\mu^\text{v}_I\boldsymbol{1}+\mu^\text{v}_D\nn \nn\right]
\label{eq:mu_v}
\end{eqnarray}
and the response of the orientation vector of the dumbbell to the local chemical gradient
\begin{eqnarray}
&&\mu^\omega \equiv \frac{\kB T}{\eta} \Bigg[ \moy{\frac{1}{l}} (A_2 - A_1)\\
&& + \frac{1}{3}\moy{\frac{1}{ l^4}} \left(a_2^3 B_2 - a_1^3 B_1 + \frac{3}{2} ( a_1^3 A_2 - a_2^3 A_1 ) \right) \Bigg]\nonumber.
\label{eq:mu_omega}
\end{eqnarray}
From Eq.~(\ref{eq:mu_v}) the translational drift velocity comes from two distinct responses to the substrate field: Diffusiophoresis of the enzyme along the substrate concentration gradient, which is controlled by $\mu^\text{v}_I$, and an anisotropic response instantaneously along $\nn$, which is controlled by $\mu^\text{v}_D$, that leads to an average drift along the gradient after a time that is larger than the rotational diffusion time. The two mechanisms have been described for active colloids, where the control parameters of the response to the chemical gradient are determined by the surface patterning of the colloid, and have both passive and active contributions \cite{saha_2014}.

The angular velocity of the enzyme will turn it towards or away from the chemical field, depending on the value of $\mu^\omega$, and hence the sign of the interactions $\phi^{is}$. This parameter will later be discussed in greater detail. 
\subsection{Polarisation}
We define the global density $c(\RR;t)=\int \text{d}\nn\, \tilde{\rho}_\text{e}$ and polar order parameter $\pp(\RR;t)=\int \text{d}\nn\, \nn\tilde{\rho}_\text{e}$ of the enzyme, which are the relevant quantities in the large time limit if we assume that rotational diffusion is fast. The equation for the polar order parameter is the first moment of (\ref{eq:smolcom}). However, the equation for $\pp$ is coupled to higher order moments of the distribution in another BBGKY hierachy. The usual protocol for truncating such a hierachy is to assume that $\pp$ is constant over sufficiently large distances and that it relaxes quickly. Additionally, that the orientation decorrelates after a sufficiently long time, so that we can set $\boldsymbol{Q}(\RR;t)=\int \text{d}\nn \left(\nn\nn-\frac{\boldsymbol{1}}{3}\right)\tilde{\rho}_\text{e}$, and higher order moments to zero. With these approximations we find the following expression for the polar order parameter
\begin{eqnarray}
\pp = \frac{1}{3D^r}\left[-D^c\nabla_{\RR}c+\mu^\omega(\nabla_{\RR}\rho_\text{s}^\infty)c\right].
\label{eq:polarity}
\end{eqnarray}
This shows that there are two ways a modular molecule can be polarised: by a density gradient through a purely hydrodynamic term which is controlled by the translation-rotation coupling scalar $D^c=\frac{\kB T}{2} \moy{\frac{\Gamma_I}{l}}$. This tends to align the enzyme so that the smaller subunit is in the lower density region and the larger subunit in the higher density region. This is because $D^c$ is positive when subunit 2 is smaller than subunit 1 and negative when subunit 2 is larger than subunit 1. The second term results in alignment due to the chemical gradient. The sign of $\mu^\omega$ is determined by the sign of the interactions between each of the subunits and the substrate molecules in the bulk. Beyond the closure approximation, there is a non-linear coupling of the two alignment mechanisms.
\section{\label{sec:mobility}Effective mobility}
Substituting the first moment of Eq.~(\ref{eq:smolcom}) into the zeroth moment, and using the moment closure scheme described above to truncate the hierachy, we derive a diffusion equation $\partial_t c(\RR;t) = D^\text{eff}\nabla^2_{\RR}c - \mu^\text{eff}\nabla_{\RR}\cdot\left[(\nabla_{\RR}\rho^\infty_\text{s}) c \right]$, with effective diffusion coeffcient and phoretic mobility
\begin{equation}
D^\text{eff} = D^t_\text{ave}-\frac{2(D^c)^2}{3D^r}
\label{eq:D^eff}
\end{equation}
and
\begin{eqnarray}
&&\mu^\text{eff} = \frac{\kB T}{\eta}\frac{\left(A_1+A_2\right)}{2}-
\frac{\kB T}{3\eta}\frac{\moy{\Gamma_I/l}}{\moy{W_I/l^2}}\Bigg[\moy{\frac{1}{l}}\left(A_2-A_1\right)\nonumber\\
&&+\frac{1}{3}\moy{\frac{1}{l^4}}\left(a_2^3B_2-a_1^3B_1+\frac{3}{2}\left(a_1^3A_2-a_2^3A_1\right)\right)\Bigg]\nonumber\\
&&=\mu^v_I+\frac{1}{3}\mu^v_D-\frac{2D^c}{D^r}\mu^\omega.
\label{eq:mu^eff}
\end{eqnarray}
The effective diffusion coefficient is as in \cite{illien_2017_EPL}, where $D^t_\text{ave}$ is the average of the translational motion and there is a negative correction due to hydrodynamic coupling of the subunits of the enzyme. At leading order, the effective phoretic mobility is the average of the Anderson-type contribution to the diffusiophoresis of the subunits, with a correction that is also due to intramolecular hydrodynamic interactions.

\begin{figure*}[t]
  \centering
  \subcaptionbox*{}[.45\linewidth][c]{%
    \includegraphics[width=.4\linewidth]{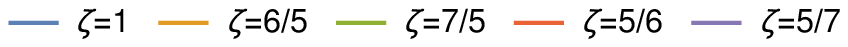}}
    \qquad
    
    \bigskip
    
  \subcaptionbox{}[.45\linewidth][c]{%
    \includegraphics[width=.4\linewidth]{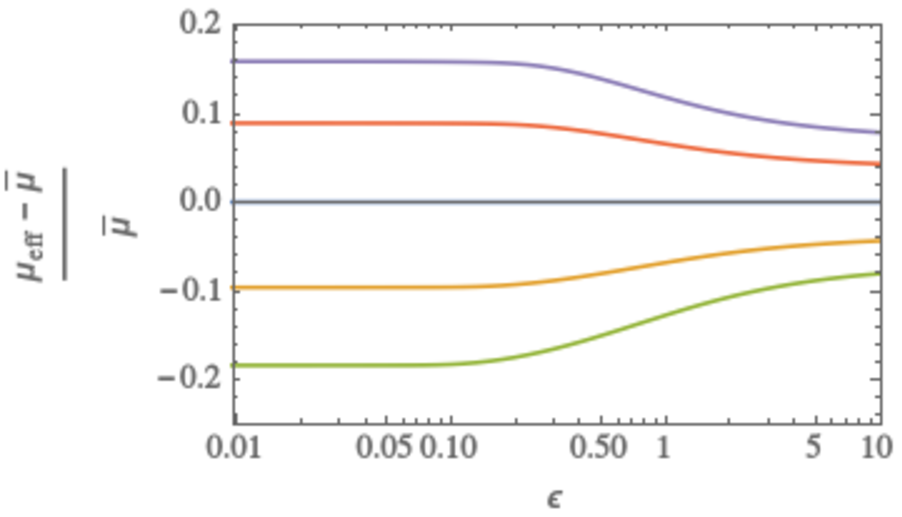}}\quad
  \subcaptionbox{}[.45\linewidth][c]{%
    \includegraphics[width=.4\linewidth]{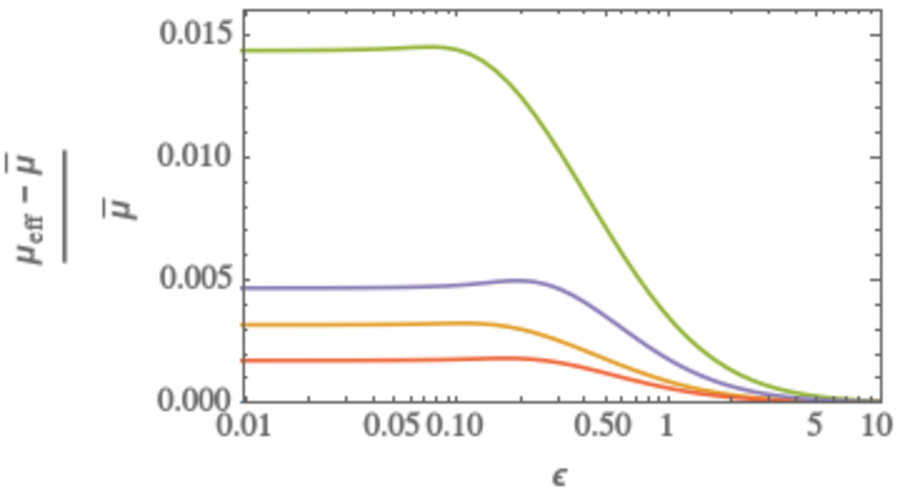}}\quad

  \bigskip

  \subcaptionbox{}[.45\linewidth][c]{%
    \includegraphics[width=.4\linewidth]{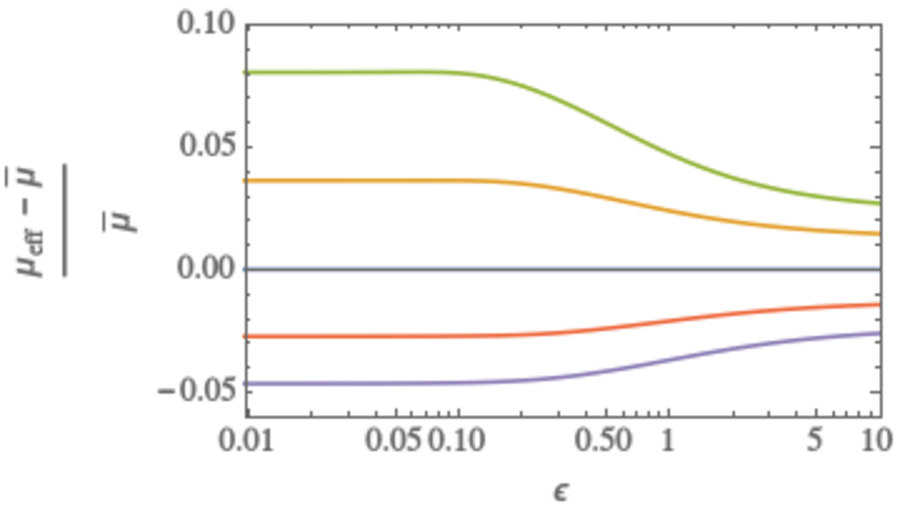}}\quad
  \subcaptionbox{}[.45\linewidth][c]{%
    \includegraphics[width=.4\linewidth]{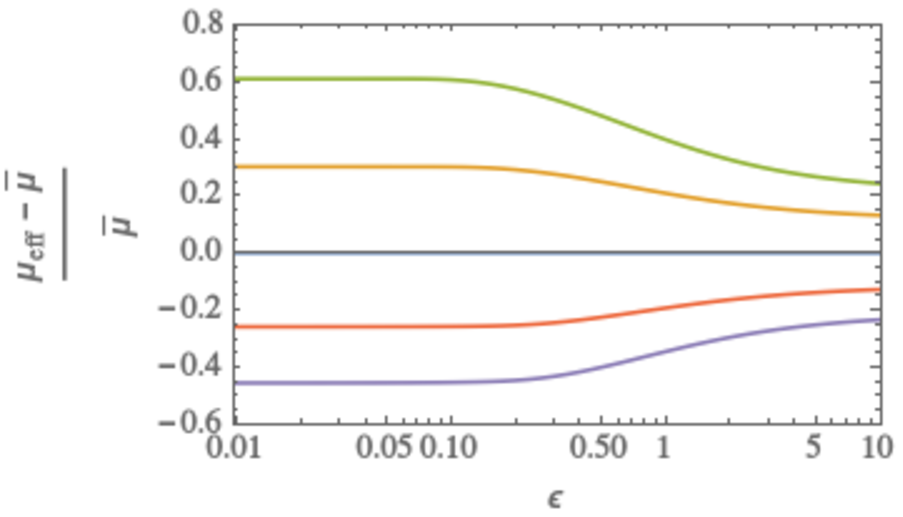}}\quad
\caption{$(\mu_\text{eff}-\bar{\mu})/\bar{\mu}$ from Eq.~(\ref{eq:muadim_ii}) for subunits interacting via a harmonic potential with stiffness $k$ as a function of the fluctuation parameter $\epsilon=\sqrt{\kB T/ka^2}$, where $a$ is the typical size of the dumbbell. The lines represent different values of the ratio between the subunit sizes $\zeta$ and the ratio $a_1/a=0.3$ is constant in all plots. The Oseen mobility functions have been used for all plots. In (a) $\tilde{A}=0$, (b) $\tilde{A}=1$, (c) $\tilde{A}=2$, and (d) $\tilde{A}=-2$. The effective mobility is increased by opposite interactions of the subunits with substrate molecules.}
   \label{fig:muadim_ii}
\end{figure*}

Defining the average mobility $\bar{\mu}=\kB T(A_1+A_2)/2\eta
$, the difference between the effective mobility of two interacting subunits and the average mobility of two non-interacting subunits is given by
\begin{eqnarray}
&&\frac{\mu_\text{eff}-\bar{\mu}}{\bar{\mu}} = -\frac{2D^c}{3D^r(1+\tilde{A})}\Bigg\{(\tilde{A}-1)\moy{\frac{1}{l}}+\nonumber\\
&&\frac{a_1^3}{3}\moy{\frac{1}{l^4}}\left[\alpha\left(\zeta^3\tilde{A}-1\right)+\frac{3}{2}\left(\tilde{A}-\zeta^3\right)\right]\Bigg\},
\label{eq:muadim_ii}
\end{eqnarray}
with $\tilde{A}=A_2/A_1$, $\alpha=B_1/A_1$ and $\zeta=a_2/a_1$. From Eqs.~(\ref{eq:Ai}) and (\ref{eq:Bi}), $\alpha$ is of the order of the particle size $a_1$ divided by the interaction length $\lambda_1$, which is about one order of magnitude for nm size enzymes and angstrom Derjaguin length. In Fig.~\ref{fig:muadim_ii} we have plotted (\ref{eq:muadim_ii}) as a function of the relative fluctuations of the enzyme for some selected values of $\tilde{A}$.

\begin{figure}[h!]
\centering
\includegraphics[width=0.4\textwidth]{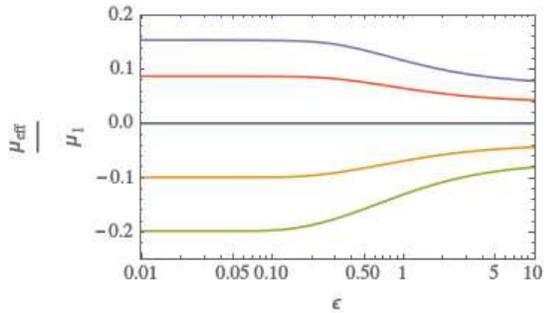}
  \caption{$\mu_\text{eff}/\mu_1$ from Eq.~(\ref{eq:muadim_i}) for subunits with balancing mobilities $A_2=-A_1$ interacting via a harmonic potential as a function of the dimensionless fluctuation parameter $\epsilon=\sqrt{\kB T/ka^2}$. $\mu_\text{eff}/\mu_1$ vanishes for subunits of equal size, and its sign is determined by the sign of the mobility of the larger subunit.}
  \label{fig:mueff_mu1}
\end{figure}

If the two subunits have balancing mobilities (i.e. $A_2=-A_1$) we can define the dimensionless quantity
\begin{eqnarray}
\frac{\mu_\text{eff}}{\mu_1} &=&\frac{2D^c}{D^r}\left[2\moy{\frac{1}{l}}+\frac{a_1^3}{3}\moy{\frac{1}{l^4}}\left(\alpha+\frac{3}{2}\right)\left(1+\zeta^3\right)\right],\nonumber\\
\label{eq:muadim_i}
\end{eqnarray}
with $\mu_1=\kB TA_1/\eta$, which shows that the sign of the effective mobility will be dictated by the sign of the mobility of the larger subunit. In Fig.~\ref{fig:mueff_mu1} we have plotted (\ref{eq:muadim_i}) as a function of the relative fluctuations of the enzyme.
\section{\label{sec:kinetics}Enzyme kinetics}
The results from the previous section can be extended to include a simplified kinetics for the enzyme. Enzymatic activity is considered as a two-state dynamical process with binding and unbinding of a substrate molecule to the enzyme binding site, but excluding the catalytic step in which the substrate is converted into product. The justification is that catalysis is typically much slower than the rate of binding activity. By introducing binding and unbinding, we find that the transport properties of the enzyme, which were predicted to inherit a space dependence and the Michaelis-Menten kinetics of the mechanochemical cycle \cite{agudo-Canalejo_2018}, attain fluctuation-induced contributions in a generic way due to the non-vanishing coupling of translational and rotational modes of the dumbbell subunits. The derivation of the effective diffusion coefficient and drift velocities presented in this section follows what is presented in Ref.~\citen{agudo-Canalejo_2018}, where an enzyme is modelled as single particle.

Assuming that the rotational diffusion time of the enzyme is faster than the mean binding time (which is typically the case for most enzymes), at a time of the order of the mean binding time, the enzyme can be either free or bound to a substrate. If $\text{S}_\text{on}(\RR,\XX)$ is the probability that an enzyme at position $\RR$ binds to a substrate molecule at position $\XX$ to form a complex at $\RR$, and $\text{S}_\text{off}(\RR,\XX)$ is the probability that the complex at $\RR$ decomposes into a free enzyme at position $\RR$ and a substrate molecule at $\XX$, the densities of the two states obey the following equations:
\begin{eqnarray}
\label{c_f1}
\partial_t c_\text{f}(\RR;t) &&= \nabla_{\RR}\cdot\left[D^\text{eff}_\text{f}\nabla_{\RR}c_\text{f} - \mu_\text{f}^\text{eff}(\nabla_{\RR}\rho^\infty_\text{s})c_\text{f} \right]\nonumber\\
&&-c_\text{f}\int_{\XX}\text{S}_\text{on}(\RR,\XX)\rho_\text{s}(\XX)\ex{-\frac{\phi^{1s}_f+\phi^{2s}_f}{\kB T}}\nonumber\\
&&+c_\text{b}\int_{\XX}\text{S}_\text{off}(\RR,\XX)
\end{eqnarray}
and
\begin{eqnarray}
\label{c_b1}
\partial_t c_\text{b}(\RR;t) &&= \nabla_{\RR}\cdot\left[D_\text{b}^\text{eff}\nabla_{\RR}c_\text{b} - \mu_\text{b}^\text{eff}(\nabla_{\RR}\rho^\infty_\text{s})c_\text{b} \right]\nonumber\\
&&+c_\text{f}\int_{\XX}\text{S}_\text{on}(\RR,\XX)\rho_\text{s}(\XX)\ex{-\frac{\phi^{1s}_b+\phi^{2s}_b}{\kB T}}\nonumber\\
&&-c_\text{b}\int_{\XX}\text{S}_\text{off}(\RR,\XX)
\end{eqnarray}
where $c_\text{f}(\RR;t)=\int \text{d}\nn \, \tilde{\rho}_\text{f}$ and $c_\text{b}(\RR;t)=\int \text{d}\nn \, \tilde{\rho}_\text{b}$ are the free and bound state densities respectively. The integrals over the substrate position $\XX$ can be performed by invoking the approximation of very short-ranged interactions to choose $\text{S}_\text{on}(\RR,\XX)=k_\text{on}\delta(\RR-\XX)$ and $\text{S}_\text{off}(\RR,\XX)=k_\text{off}\delta(\RR-\XX)$, where $k_\text{on}$ and $k_\text{off}$ are the association and decomposition rates. After redefining $k_\text{on}$ to include the constant $\exp{[-(\phi^{1s}_f(0)+\phi^{2s}_f(0))/\kB T]}$, (\ref{c_f1}) and (\ref{c_b1}) become
\begin{eqnarray}
\label{c_f2}
\partial_t c_\text{f}(\RR;t) &&= \nabla_{\RR}\cdot\left[D_\text{f}^\text{eff}\cdot\nabla_{\RR}c_\text{f} - \mu_\text{f}^\text{eff}(\nabla_{\RR}\rho^\infty_\text{s})c_\text{f} \right]\nonumber\\
&&-k_\text{on}\rho_\text{s}(\RR)c_\text{f}+k_\text{off}c_\text{b}
\end{eqnarray}
and
\begin{eqnarray}
\label{c_b2}
\partial_t c_\text{b}(\RR;t) &&= \nabla_{\RR}\cdot\left[D_\text{b}^\text{eff}\cdot\nabla_{\RR}c_\text{b} - \mu_\text{b}^\text{eff}(\nabla_{\RR}\rho^\infty_s)c_\text{b} \right]\nonumber\\
&&+k_\text{on}\rho_\text{s}(\RR)c_\text{f}-k_\text{off}c_\text{b}.
\end{eqnarray}
The experimentally relevant quantity is the total enzyme concentration   
\begin{equation}
c_\text{tot}(\RR;t)=c_\text{f}(\RR;t)+c_\text{b}(\RR;t).
\label{total}
\end{equation}
In the experimental set-up the free enzyme is indistinguishable from the enzyme-substrate complex. Furthermore, we can assume that binding is at local and instantaneous equilibrium since the time for motion in $\RR$ is much greater than the binding time. This gives
\begin{equation}
k_\text{on}c_\text{f}(\RR;t)\rho_\text{s}(\RR;t)\approx k_\text{off}c_\text{b}(\RR;t)
\end{equation}
and
\begin{eqnarray}
c_\text{f}=\frac{K}{K+\rho_\text{s}}c_\text{tot} \quad \quad  c_\text{b}=\frac{\rho_\text{s}}{K+\rho_\text{s}}c_\text{tot}.
\label{eq:cf_cb}
\end{eqnarray}
Finally, the sum of (\ref{c_f2}) and (\ref{c_b2}) using (\ref{eq:cf_cb}) gives Eq.~(\ref{eq:c_tot}) for the total enzyme concentration, where the transport coefficients, including the drift velocity due to the difference in diffusion coefficient of the free and bound state enzyme \cite{agudo-Canalejo_2018}, are averaged over the two states with a Michaelis-Menten weight. Specifically,
\begin{equation}
D^\text{eff}(\RR) = D_\text{f}^\text{eff}+(D_\text{b}^\text{eff}-D_\text{f}^\text{eff})\frac{\rho_\text{s}(\RR)}{K+\rho_\text{s}(\RR)},
\label{eq:D_binding}
\end{equation}
\begin{equation}
\label{eq:V_ph_binding}
\VV^\text{eff}(\RR) = \VV^\text{eff}_\text{f}(\RR)+[\VV^\text{eff}_\text{b}(\RR)-\VV^\text{eff}_\text{f}(\RR)]\frac{\rho_\text{s}(\RR)}{K+\rho_\text{s}(\RR)},
\end{equation}
and
\begin{equation}
\label{eq:V_bi_binding}
\VV^\text{eff}_\text{bi}(\RR) = -(D_\text{b}^\text{eff}-D_\text{f}^\text{eff})\nabla_{\RR}\left(\frac{\rho_\text{s}(\RR)}{K+\rho_\text{s}(\RR)}\right),
\end{equation}
where $D^\text{eff}_\text{i}$ is given by (\ref{eq:D^eff}) and $\VV^\text{eff}_\text{i}(\RR)=\mu^\text{eff}_\text{i}\nabla_{\RR}\rho_\text{s}(\RR)$, with $\mu^\text{eff}_\text{i}$ given by (\ref{eq:mu^eff}). $\VV^\text{eff}_\text{bi}(\RR)=-\nabla_{\RR}D^\text{eff}(\RR)$, and in the absence of phoresis Eq.~(\ref{eq:c_tot}) can be written as $
\partial_t c_\text{tot}(\RR;t) = \nabla_{\RR}^2\left\{D^\text{eff}(\RR)c_\text{tot}\right\}$

The form of Eq.~(\ref{eq:c_tot}) depends on details of the enzymes' kinetics. We have focused on the case of rotational diffusion being faster than the mean binding time. If instead one assumes that the binding rate is much faster than rotational diffusion, the hydrodynamic corrections in (\ref{eq:D_binding})-(\ref{eq:V_bi_binding}) will have different forms. Furthermore, $\VV^\text{eff}_\text{bi}$ can no longer be written exactly as the derivative of the diffusion coefficient. Generically, any quantity that depends on the coupling between the fast (local) and slow (global) dynamics will be determined by this time-scale separation.
\section{\label{conclusions} Conclusions}
The specific geometry of an asymmetric dumbbell was used to study the response of an enzyme to an inhomogeneous concentration of its substrate in order to access the hydrodynamic interactions that arise in the flexible, modular structure of an enzyme. The interactions of the enzyme with the concentration gradient produces a drift velocity and a tendency to align parallel or antiparallel to the gradient, depending on the sign of the interactions between the enzyme and substrate molecules. We find a second alignment mechanism, due to gradients in the density field of the enzyme, that is controlled by the strength of the coupling between the translational and rotational motion of the enzyme. The effects of the two alignment mechanisms, particularly of alignment due to hydrodynamic interactions, and the non-linear coupling of the two are likely to be significant for the collective behaviour of many interacting enzyme molecules.

In previous work, we showed that the diffusion coefficient of a modular macromolecule is overestimated if it is considered as a rigid, symmetric object. This is shown to be a generic feature that is also true of the drift veclocity. Though the form of the fluctuation-induced corrections we present is specific to the time-scale ordering and hence a chosen enzyme kinetics, the case we considered is the most relevant given that it corresponds to the ordering that is most observed.

Finally, we note that our analysis and results can be easily generalised, with further generalisations of our model or others that contain similar interactions and geometric considerations within the low-Reynolds number framework. 
\begin{acknowledgments}
We thank Pierre Illien and Suropriya Saha for invaluable discussions. This work was supported by the U.S. National Science Foundation
under MRSEC grant no. DMR-1420620. T.A-L. acknowledges the support of an EPSRC Studentship.
\end{acknowledgments}

\appendix
\section{\label{app:phoretic_integration} Integration of phoretic terms}
\subsection{Self-type phoretic terms}
Integrals of the type
\begin{equation}
\int_{\XX}(\boldsymbol{\mu}^{is} - \boldsymbol{\mu}^{ii}) ( \ex{-\frac{\phi^{is}}{\kB T}} - 1) \nabla_{\XX} \rho_s
\label{intA}
\end{equation}
are identical to those appearing in the case of phoresis of a single particle.

For a spherical subunit $i$ of radius $a_i$, assuming the substrate molecules are point-like, we have the mobility tensors
\begin{equation}
\boldsymbol{\mu}^{ii} = \frac{1}{6 \pi \eta a_i} \boldsymbol{1}
\end{equation}
and
\begin{equation}
\boldsymbol{\mu}^{is} = \frac{1}{6 \pi \eta a_i} \left[ \frac{1}{4} \left( 3 \frac{a_i}{r_i} + \frac{a_i^3}{r_i^3} \right) \boldsymbol{1} + \frac{3}{4} \left(\frac{a_i}{r_i} - \frac{a_i^3}{r_i^3} \right) \boldsymbol{\hat{r}_i}\boldsymbol{\hat{r}_i} \right],
\end{equation}
where $r_i$ is the distance between the center of subunit $i$ and the substrate molecule, and $\boldsymbol{\hat{r}_i}$ is the radial unit vector. Combining both, we have
\begin{eqnarray}
\boldsymbol{\mu}^{is} - \boldsymbol{\mu}^{ii} &&= \frac{1}{6 \pi \eta a_i} \Bigg[\left( -1 + \frac{3}{4} \frac{a_i}{r_i} + \frac{1}{4} \frac{a_i^3}{r_i^3} \right) (\boldsymbol{1}-\boldsymbol{\hat{r}_i}\boldsymbol{\hat{r}_i})\nonumber\\
&& + \left(-1 + \frac{3}{2} \frac{a_i}{r_i} - \frac{1}{2}\frac{a_i^3}{r_i^3} \right) \boldsymbol{\hat{r}_i}\boldsymbol{\hat{r}_i} \Bigg].
\end{eqnarray}

The concentration of substrate in the proximity of a subunit, assuming that far enough from the subunit there is a concentration gradient $\nabla \rho_s^\infty = |\nabla \rho_s^\infty| \boldsymbol{\hat{z}}$ of substrate molecules pointing in the $z$-direction, is given by the solution to the Laplace quation with no normal flux boundary conditions on the surface of the sphere
\begin{equation}
\rho_s (r_i,\theta_i,\varphi_i) = A + |\nabla \rho_s^\infty| \left( r_i + \frac{1}{2} \frac{a_i^3}{r_i^2} \right) \cos \theta_i,
\end{equation}
where we use spherical coordinates centred on the subunit $i$. Using the gradient operator in spherical coordinates $\nabla f = \partial_r f \boldsymbol{\hat{r}} + (1/r) \partial_\theta f \boldsymbol{\hat{\theta}} + (1/r \sin \theta) \partial_\varphi f \boldsymbol{\hat{\varphi}}$, we calculate
\begin{equation}
\nabla \rho_s  =  |\nabla \rho_s^\infty| \left[ \left( 1 - \frac{a_i^3}{r_i^3} \right) \cos \theta_i \boldsymbol{\hat{r}_i} - \left( 1 + \frac{1}{2} \frac{a_i^3}{r_i^3} \right) \sin \theta_i \boldsymbol{\hat{\theta}_i}   \right].
\label{eq:laplace}
\end{equation}

In order to evaluate (\ref{intA}), we need
\begin{equation}
(\boldsymbol{1}-\boldsymbol{\hat{r}_i}\boldsymbol{\hat{r}_i}) \nabla \rho_s = - |\nabla \rho_s^\infty| \left( 1 + \frac{1}{2} \frac{a_i^3}{r_i^3} \right) \sin \theta_i \boldsymbol{\hat{\theta}_i}
\end{equation}
and
\begin{equation}
(\boldsymbol{\hat{r}_i}\boldsymbol{\hat{r}_i}) \nabla \rho_s = |\nabla \rho_s^\infty|  \left( 1 - \frac{a_i^3}{r_i^3} \right) \cos \theta_i \boldsymbol{\hat{r}_i}.
\end{equation}
Furthermore, using the definition in cartesian coordinates of $\boldsymbol{\hat{r}} = \sin \theta \cos \varphi \boldsymbol{\hat{x}} + \sin \theta \sin \varphi \boldsymbol{\hat{y}} + \cos \theta  \boldsymbol{\hat{z}}$ and $\boldsymbol{\hat{\theta}} = \cos \theta \cos \varphi \boldsymbol{\hat{x}} + \cos \theta \sin \varphi \boldsymbol{\hat{y}} - \sin \theta  \boldsymbol{\hat{z}}$, we can calculate the integrals over the solid angle $\int \mathrm{d}\Omega \sin \theta \boldsymbol{\hat{\theta}} = -(8\pi/3) \boldsymbol{\hat{z}}$ and $\int \mathrm{d}\Omega \cos \theta \boldsymbol{\hat{r}} = (4\pi/3) \boldsymbol{\hat{z}}$.

Taking together all these results, and noting that the two sets of coordinates used are related to each other by $\XX = \RR + r_i \boldsymbol{\hat{r}_i}$, we can finally evaluate (\ref{intA}) to be
\begin{equation}
\int_{\XX}(\boldsymbol{\mu}^{is} - \boldsymbol{\mu}^{ii}) ( \ex{-\frac{\phi^{is}}{\kB T}} - 1) \nabla_{\XX} \rho_s= - \frac{A_i}{\eta} \nabla_{\RR} \rho_s^\infty,
\label{eq:anderson}
\end{equation}
where we have defined
\begin{equation}
A_i \equiv \frac{1}{6 a_i} \int^\infty_{a_i}\text{d}r_i\,r^2_i\,(\ex{-\frac{\phi^{is}}{\kB T}}-1)\left(4-4\frac{a_i}{r_i}+\frac{a_i^4}{r_i^4}-\frac{a_i^6}{r_i^6}\right).
\label{Aig}
\end{equation}
For very short ranged interactions we can use the approximation $r_i = a_i + \delta$ with $\delta \ll a_i$. The terms inside the rightmost parenthesis in the integral become $6 \delta/a_i$ to lowest order, giving (\ref{eq:Ai}) from the main text.
\subsection{Cross-type phoretic terms}
We also need to evaluate integrals of the type
\begin{equation}
\int_{\XX}(\boldsymbol{\mu}^{is} - \boldsymbol{\mu}^{12}) ( \ex{-\frac{\phi^{js}}{\kB T}} - 1) \nabla_{\XX} \rho_s,
\label{intB}
\end{equation}
with $i\neq  j$.

For spherical subunits of radius $a_i$, and point-like substrate molecules, the mobility tensors are
\begin{eqnarray}
\boldsymbol{\mu}^{is} &=& \frac{1}{6\pi \eta a_i}\left[\frac{1}{4}\left(3\frac{a_i}{r_i}+\frac{a^3_i}{r^3_i}\right)\boldsymbol{1}+\frac{3}{4}\left(\frac{a_i}{r_i}-\frac{a^3_i}{r^3_i}\right)\rr_i\rr_i\right]\nonumber\\
\end{eqnarray}
and
\begin{eqnarray}
\muOT &\simeq & \frac{1}{8\pi \eta}\left[\left(\frac{1}{l}+\frac{a^2_1+a^2_2}{3l^3}\right)\boldsymbol{1}+\left(\frac{1}{l}-\frac{a^2_1+a^2_2}{l^3}\right)\nn\nn\right].\nonumber\\
\end{eqnarray}

In order to evaluate (\ref{intB}), we are interested in evaluating $\boldsymbol{\mu}^{2s}$ in the proximity of particle $1$, and $\boldsymbol{\mu}^{1s}$ in the proximity of particle $2$. Using the relation $\boldsymbol{r}_2 =  \boldsymbol{r}_1 - \boldsymbol{l}$, and expanding in powers of $r_1/l$ up to order $O(r_1^2/l^2) \sim O(a_1^2/l^2)$, which is of the same order as $\muOT$, we can write $\boldsymbol{\mu}^{2s}$ as
\begin{eqnarray}
\boldsymbol{\mu}^{2s} & \simeq & \frac{1}{8\pi \eta}\bigg\{\left[\frac{1}{l}\left(1+\alpha_1 \frac{r_1}{l} + \frac{1}{2} (3\alpha_1^2-1)\frac{r_1^2}{l^2} \right)+\frac{1}{3}\frac{a_2^2}{l^3}\right]\boldsymbol{1}\nonumber\\
&+&\left[\frac{1}{l} \left( 1 + 3 \alpha_1 \frac{r_1}{l} + \frac{3}{2} (5 \alpha_1^2 -1) \frac{r_1^2}{l^2} \right)-\frac{a^2_2}{l^3}\right]\nn\nn \nonumber \\
&-& \frac{r_1}{l^2} \left[ 1 + 3 \alpha_1 \frac{r_1}{l} \right] (\rr_1 \nn + \nn \rr_1) + \frac{r_1^2}{l^3} \rr_1 \rr_1 \bigg\},
\end{eqnarray}
with $\alpha_1 \equiv \nn \cdot \rr_1$. A similar expression can be obtained for $\boldsymbol{\mu}^{1s}$ in the proximity of particle 2, with
\begin{eqnarray}
\boldsymbol{\mu}^{1s} & \simeq & \frac{1}{8\pi \eta}\bigg\{\left[\frac{1}{l}\left(1-\alpha_2 \frac{r_2}{l} + \frac{1}{2} (3\alpha_2^2-1)\frac{r_2^2}{l^2} \right)+\frac{1}{3}\frac{a_1^2}{l^3}\right]\boldsymbol{1}\nonumber\\
&+&\left[\frac{1}{l} \left( 1 - 3 \alpha_2 \frac{r_2}{l} + \frac{3}{2} (5 \alpha_2^2 -1) \frac{r_2^2}{l^2} \right)-\frac{a^2_1}{l^3}\right]\nn\nn \nonumber \\
&+& \frac{r_2}{l^2} \left[ 1 - 3 \alpha_2 \frac{r_2}{l} \right] (\rr_2 \nn + \nn \rr_2) + \frac{r_2^2}{l^3} \rr_2 \rr_2 \bigg\},
\end{eqnarray} 
with $\alpha_2 \equiv \nn \cdot \rr_2$.

By considering the concentration of substrate centered around each subunit, and assuming that far from each subunit there is a concentration gradient $\nabla_{\RR}\rho^\infty_s=|\nabla_{\RR}\rho^\infty_s|\hat{z}$ of substrate molecules in the $z$-direction, the concentration gradient can be extracted from the integral in the same manner as for the self-type phoretic term above, giving
\begin{eqnarray}
&&\int_{\XX}(\boldsymbol{\mu}^{is} - \boldsymbol{\mu}^{12}) ( \ex{-\frac{\phi^{js}}{\kB T}} - 1) \nabla_{\XX} \rho_s\nonumber\\
&&= - \frac{a_j^3}{l^3} \frac{B_j}{\eta} \left(\nn\nn-\frac{\boldsymbol{1}}{3}\right) \nabla_{\RR}\rho^\infty_s,
\label{intB2}
\end{eqnarray}
where we have defined
\begin{equation}
B_i \equiv \frac{1}{10} \int^\infty_{a_i}\text{d}r_i\,r_i\,(\ex{-\frac{\phi^{is}}{\kB T}}-1) \left( 1 - 5 \frac{r_i}{a_i} + 5 \frac{r_i^3}{a_i^3} \right).
\label{Big}
\end{equation}
In this case, considering very short ranged interactions we find (\ref{eq:Bi}).
\subsection{Higher order corrections due to perturbation of solute concentration field by the second subunit}
In the two preceding subsections, we have assumed that the concentration of substrate molecules around any given subunit is unaffected by the presence of the other subunit. Here we show that this is not the case, and calculate the lowest order correction to the concentration field, and its effect on the phoretic terms.

If we did a naive superposition of the fields around two particles of radii $a_1$ and $a_2$, we would write it as
\begin{equation}
\nabla \rho_s^\text{naive}  =  \nabla\rho_s^{(0)} + \nabla\rho_s^{(1)} + \nabla\rho_s^{(2)},
\end{equation}
with
\begin{eqnarray}
\label{0}
\nabla \rho_s^{(0)} &&=  |\nabla \rho_s^\infty| \left[ \cos \theta_1 \boldsymbol{\hat{r}}_1 - \sin \theta_1 \boldsymbol{\hat{\theta}}_1   \right]\\
&& = |\nabla \rho_s^\infty| \left[ \cos \theta_2 \boldsymbol{\hat{r}}_2 - \sin \theta_2 \boldsymbol{\hat{\theta}}_2   \right] = |\nabla \rho_s^\infty| \boldsymbol{\hat{z}}\nonumber,
\end{eqnarray}
\begin{eqnarray}
\nabla \rho_s^{(1)}  &&=  - |\nabla \rho_s^\infty| \frac{a_1^3}{r_1^3} \left[ \cos \theta_1 \boldsymbol{\hat{r}}_1 + \frac{1}{2} \sin \theta_1 \boldsymbol{\hat{\theta}}_1   \right]\nonumber\\
&&= |\nabla \rho_s^\infty| \frac{a_1^3}{r_1^3} \left[ - \boldsymbol{\hat{r}}_1 \boldsymbol{\hat{r}}_1 + \frac{1}{2} \boldsymbol{\hat{\theta}}_1  \boldsymbol{\hat{\theta}}_1   \right] \boldsymbol{\hat{z}}
\label{1}
\end{eqnarray}
and
\begin{eqnarray}
\nabla \rho_s^{(2)}  &&=  - |\nabla \rho_s^\infty| \frac{a_2^3}{r_2^3} \left[ \cos \theta_2 \boldsymbol{\hat{r}}_2 + \frac{1}{2} \sin \theta_2 \boldsymbol{\hat{\theta}}_2   \right]\nonumber\\
&&= |\nabla \rho_s^\infty| \frac{a_2^3}{r_2^3} \left[ - \boldsymbol{\hat{r}}_2 \boldsymbol{\hat{r}}_2 + \frac{1}{2} \boldsymbol{\hat{\theta}}_2  \boldsymbol{\hat{\theta}}_2   \right] \boldsymbol{\hat{z}}
\label{2}
\end{eqnarray}
using Eq.~(\ref{eq:laplace}) for the solution around a single subunit, where $\boldsymbol{\hat{r}}_i, \boldsymbol{\hat{\theta}}_i$ are the unit vectors of spherical coordinates centered at the $i$-th subunit, both sharing $\boldsymbol{\hat{z}}$ as the zenith direction. The two sets of coordinates are related by $r_1 \rr_1 = l \nn + r_2 \rr_2$.

The naive superposition is a solution of the Laplace equation, and it satisfies the boundary condition at infinity, but it does \emph{not} satisfy the no normal flux boundary condition at the surface of the subunits. Indeed, the contribution $\nabla \rho_s^{(2)}$ in the proximity of subunit 1 is, to lowest order in $1/l$,
\begin{eqnarray}
\label{2prox}
&&\nabla \rho_s^{(2)} =  \nabla \rho_s^{(2)}(r_1=0) + O \left( |\nabla \rho_s^\infty| \frac{a_2^3 r_1}{l^4} \right) \\
&&= |\nabla \rho_s^\infty| \frac{3}{2} \frac{a_2^3}{l^3} \left[ \frac{\boldsymbol{1}}{3} - \nn \nn  \right] \boldsymbol{\hat{z}} +  O \left( |\nabla \rho_s^\infty| \frac{a_2^3 r_1}{l^4} \right)\nonumber,
\end{eqnarray}
where we have used the fact that at the center of subunit 1 we have $r_2 = l$, $\boldsymbol{\hat{r}}_2 \boldsymbol{\hat{r}}_2 \cdot \boldsymbol{\hat{z}} = \nn \nn \cdot \boldsymbol{\hat{z}}$, and $\boldsymbol{\hat{\theta}}_2  \boldsymbol{\hat{\theta}}_2 \cdot \boldsymbol{\hat{z}} = (\boldsymbol{1} - \nn \nn ) \cdot \boldsymbol{\hat{z}}$. Therefore, to order $1/l^3$, the effect of the presence of subunit 2 is that it generates a uniform gradient around 1, that violates the no flux boundary condition. In order to cancel out this gradient while still satisfying the Laplace equation, we need to add a new term of the form
\begin{equation}
\nabla \rho_s^{(1,n)}  =  |\nabla \rho_s^\infty| \frac{3}{2} \frac{a_2^3}{l^3} \frac{a_1^3}{r_1^3} \left[ - \boldsymbol{\hat{r}}_1 \boldsymbol{\hat{r}}_1 + \frac{1}{2} \boldsymbol{\hat{\theta}}'_1  \boldsymbol{\hat{\theta}}'_1   \right] \left[ \frac{\boldsymbol{1}}{3} - \nn \nn  \right] \boldsymbol{\hat{z}}
\label{1n}
\end{equation}
where the unit vector $\boldsymbol{\hat{\theta}}'_1$ corresponds to spherical coordinates centred at subunit 1, but with the zenith direction parallel to $\left[\boldsymbol{1}/3 - \nn \nn  \right] \boldsymbol{\hat{z}}$. It is easy to see that (\ref{1n}) will cancel out the contribution of (\ref{2prox}) to the radial gradient at the surface of subunit 1 in the exact same way that (\ref{1}) cancels out the contribution of (\ref{0}). The same argument can be applied to the gradient generated by subunit 1 in the proximity of 2, resulting in a new term of the form
\begin{equation}
\nabla \rho_s^{(2,n)}  =  |\nabla \rho_0^\infty| \frac{3}{2} \frac{a_1^3}{l^3} \frac{a_2^3}{r_2^3} \left[ - \boldsymbol{\hat{r}}_2 \boldsymbol{\hat{r}}_2 + \frac{1}{2} \boldsymbol{\hat{\theta}}'_2  \boldsymbol{\hat{\theta}}'_2   \right] \left[ \frac{\boldsymbol{1}}{3} - \nn \nn  \right] \boldsymbol{\hat{z}}
\label{2n}
\end{equation}
where the unit vector $\boldsymbol{\hat{\theta}}'_2$ corresponds to spherical coordinates centred at subunit 2 with the zenith direction parallel to $\left[\boldsymbol{1}/3 - \nn \nn  \right] \boldsymbol{\hat{z}}$.

All in all, the solute concentration gradient can therefore be written to order $a_i^3 / l^3$ as 
\begin{equation}
\nabla \rho_s \approx  \nabla\rho_s^{(0)} + \nabla\rho_s^{(1)} + \nabla\rho_s^{(2)} + \nabla\rho_s^{(1,n)} + \nabla\rho_s^{(2,n)}.
\end{equation}
This expression satisfies exactly the Laplace equation and the boundary condition at infinity, and it also satisfies the no normal flux boundary conditions at the surface of the subunits up to order $a_i^3/l^3$. 

The $a_i^3/l^3$ corrections to $\nabla \rho_s$ just described have an effect on the $A$-type contributions to the phoretic velocity. The integral (\ref{eq:anderson}) picks up a correction, becoming
\begin{eqnarray}
&&\int_{\XX}\left(\boldsymbol{\mu}^{is}-\boldsymbol{\mu}^{ii}\right)\cdot(\ex{-\frac{\phi^{is}}{\kB T}}-1)\nabla_{\XX}\rho_s \nonumber\\
&& = - \frac{A_i}{\eta} \left[ 1 + \frac{3}{2} \frac{a_j^3}{l^3} \left( \frac{\boldsymbol{1}}{3} - \nn \nn \right) \right] \nabla_{\RR} \rho_s^\infty.
\label{intA2}
\end{eqnarray}
The $B$-type contribution (\ref{intB2}) is however unchanged to order $a_i^3/l^3$, because the corresponding corrections would be of order $a_i^6/l^6$.
\section{\label{nabla_l_decomposition} Decomposition of $\nabla_{\LL}$ into separation and orientation components}

There are two ways in which one can decompose gradients and divergences of the form $\nabla_{\LL} A$ and $\nabla_{\LL}  \cdot \AAA$ into separation and orientation parts. One way is to use the orientational part of the $\nabla_{\LL}$ operator $\rotopt = \partial/\partial \nn = l (\II-\nn \nn)\nabla_{\LL}$, in which case we have
\begin{equation}
\nabla_{\LL} A = \frac{\partial A}{\partial l} \nn + \frac{1}{l} \rotopt A,
\end{equation}
with
\begin{equation}
\rotopt A = \frac{\partial A}{\partial \nn} = \frac{\partial A}{\partial \theta} \uth + \frac{1}{\sin \theta} \frac{\partial A}{\partial \varphi}\uph,
\end{equation}
as well as
\begin{equation}
\nabla_{\LL} \cdot \AAA =  \frac{1}{l^2} \frac{\partial (l^2 \AAA\cdot \nn)}{\partial l}  + \frac{1}{l} \rotopt \cdot \AAA,
\end{equation}
with
\begin{equation}
\rotopt \cdot \AAA = \frac{\partial}{\partial \nn} \cdot \AAA =  \frac{1}{\sin \theta} \frac{\partial (\sin \theta \AAA \cdot \uth)}{\partial \theta}  + \frac{1}{\sin \theta} \frac{\partial (\AAA \cdot \uph)}{\partial \varphi}.
\end{equation}
Alternatively, one can use the angular momentum operator $\rotop = \nn \times \partial/\partial \nn$, which satisfies the identities
\begin{eqnarray}
&&\rotopt A = - \nn \times (\rotop A)\nonumber,\\
&&\rotopt \cdot \AAA = \rotop \cdot ( \nn \times \AAA )\nonumber,\\
&&\rotopt \cdot (\rotopt A) = \rotop \cdot (\rotop A).
\label{identities}
\end{eqnarray}
The Smoluchowski equation (\ref{eq:smolcom}) can be decomposed in this way to give
\begin{widetext}
\begin{eqnarray}
&&\partial_t\rho_\text{e}(\RR, \nn, l; t) =  \frac{1}{4}\nabla_{\RR}\cdot(\kB T \MM\cdot\nabla_{\RR}\rho_\text{e}) +  \rotopt \cdot \left( \kB T \frac{W_I}{l^2} \rotopt \rho_\text{e} \right) + \frac{1}{2} \rotopt \cdot \left( \kB T \frac{\Gamma_I}{l} \nabla_{\RR} \rho_\text{e} \right)\nonumber \\
&&+\frac{1}{2} \nabla_{\RR} \cdot \left( \kB T \frac{\Gamma_I}{l} \rotopt \rho_\text{e} \right)+ \frac{1}{2} \nabla_{\RR} \cdot \left[ (\Gamma_I + \Gamma_D) \left( \kB T \frac{\partial \rho_\text{e}}{\partial l} + U' \rho_\text{e} \right) \nn \right]\nonumber\\
&& + \frac{1}{2} \frac{1}{l^2} \frac{\partial}{\partial l} \left[ l^2 \kB T (\Gamma_I + \Gamma_D) \nn \cdot \nabla_{\RR} \rho_\text{e} \right] +  \frac{1}{l^2} \frac{\partial}{\partial l} \left[ l^2 (W_I + W_D) \left( \kB T \frac{\partial \rho_\text{e}}{\partial l} + U' \rho_\text{e}\right) \right] \nonumber \\
&&- \nabla_{\RR} \cdot \left\{  \frac{\kB T}{2\eta} \rho_\text{e} \left[ A_1 + A_2 + \frac{1}{l^3} \left(a_1^3 B_1 + a_2^3 B_2 - \frac{3}{2} (a_2^3 A_1 + a_1^3 A_2 ) \right) \left(  \nn \nn - \frac{\boldsymbol{1}}{3} \right) \right] \cdot \nabla_{\RR} \rho_s^\infty  \right\} \nonumber \\
&&- \rotopt \cdot \left\{  \frac{\kB T}{\eta} \rho_\text{e} \left[ \frac{A_2 - A_1}{l} + \frac{1}{3} \frac{1}{ l^4} \left(a_2^3 B_2 - a_1^3 B_1 + \frac{3}{2} ( a_1^3 A_2 - a_2^3 A_1 ) \right) \right]  \nabla_{\RR} \rho_s^\infty  \right\}\nonumber\\
&& - \frac{1}{l^2} \frac{\partial}{\partial l}  \left\{ l^2 \frac{\kB T}{\eta} \rho_\text{e} \left[(\sgm_2-\sgm_1) \cdot \nabla_{\RR} \rho_s^\infty \right] \cdot \nn \right\},
\label{smolcom2}
\end{eqnarray}
\end{widetext}
which after performing the separation averaging and using the identities (\ref{identities}), we can rewrite as
\begin{widetext}
\begin{eqnarray}
&&\partial_t\tilde{\rho}_\text{e}(\RR, \nn; t) = \frac{1}{4}\nabla_{\RR}\cdot(\kB T \moy{\MM}\cdot\nabla_{\RR}\tilde{\rho}_\text{e}) +  \rotop \cdot \left( \kB T \moy{\frac{W_I}{l^2}} \rotop \tilde{\rho}_\text{e} \right)\nonumber \\
&& + \frac{1}{2} \rotop \cdot \left( \kB T \moy{\frac{\Gamma_I}{l}} (\nn \times \nabla_{\RR} \tilde{\rho}_\text{e}) \right) - \frac{1}{2} \nabla_{\RR} \cdot \left( \kB T \moy{\frac{\Gamma_I}{l}} (\nn \times \rotop \tilde{\rho}_\text{e}) \right)\nonumber\\
&&- \nabla_{\RR} \cdot \left\{  \frac{\kB T}{2\eta} \tilde{\rho}_\text{e} \left[ A_1 + A_2 + \moy{\frac{1}{l^3}} \left(a_1^3 B_1 + a_2^3 B_2 - \frac{3}{2} (a_2^3 A_1 + a_1^3 A_2 ) \right) \left(  \nn \nn - \frac{\boldsymbol{1}}{3} \right) \right] \cdot \nabla_{\RR} \rho_s^\infty  \right\} \nonumber \\
&&- \rotop \cdot \left\{  \frac{\kB T}{\eta} \tilde{\rho}_\text{e} \left[ \moy{\frac{1}{l}} (A_2 - A_1) + \frac{1}{3}\moy{\frac{1}{ l^4}} \left(a_2^3 B_2 - a_1^3 B_1 + \frac{3}{2} ( a_1^3 A_2 - a_2^3 A_1 ) \right) \right] ( \nn \times \nabla_{\RR} \rho_s^\infty) \right\},
\label{smolcom4}
\end{eqnarray}
\end{widetext}
from which Eq.~(\ref{eq:canon}) follows.
\nocite{*}
\bibliography{references}% Produces the bibliography via BibTeX.

\end{document}